\documentclass[iop,numberedappendix]{emulateapj}

\def\mdot{\hbox{$\dot {\it M}$}}
\def\micron{$\mu$m}
\def\microns{$\mu$m}
\def\lsun{\rm {\it L}_{\sun}}

\newcommand\be{\begin{equation}}
\newcommand\en{\end{equation}}

\newcounter{column_number}
\begin{document}

\shortauthors{Espaillat et al.}
\shorttitle{IRAS~04125+2902}

\title{The Transitional Disk around IRAS~04125+2902}

\author{
C. Espaillat\altaffilmark{1}, S. Andrews\altaffilmark{2}, D. Powell\altaffilmark{2}, D. Feldman\altaffilmark{1}, C. Qi\altaffilmark{2}, D. Wilner\altaffilmark{2}, \& P. D'Alessio\altaffilmark{3,4}
}

\altaffiltext{1}{Department of Astronomy, Boston University, 725 Commonwealth Avenue, Boston, MA 02215, USA; cce@bu.edu}
\altaffiltext{2}{Harvard-Smithsonian Center for Astrophysics, 60 Garden
Street, Cambridge, MA, 02138, USA; sandrews@cfa.harvard.edu, cqi@cfa.harvard.edu, dwilner@cfa.harvard.edu}
\altaffiltext{3}{Centro de Radioastronom\'{i}a y Astrof\'{i}sica,
Universidad Nacional Aut\'{o}noma de M\'{e}xico, 58089 Morelia,
Michoac\'{a}n, M\'{e}xico}
\altaffiltext{4}{Deceased Nov 14, 2013}

\begin{abstract} 

Resolved submillimeter imaging of transitional disks is increasingly
revealing the complexity of disk structure. Here we present the first
high-resolution submillimeter image 
of a recently identified
transitional disk around IRAS~04125+2902 in the Taurus
star-forming region.  
We measure an inner disk
hole of $\sim$20~AU around IRAS~04125+2902 
by simultaneously modeling new 880~{\micron} Submillimeter Array (SMA) data 
along with an existing spectral energy distribution supplemented by new Discovery Channel Telescope (DCT) photometry.
We also constrain the outer radius of the
dust disk in IRAS~04125+2902 to $\sim$50--60 AU. 
Such a small dust disk could be attributed to initial formation conditions,
outward truncation by an unseen companion, or dust evolution in the
disk. Notably, the dust distribution of IRAS~04125+2902 resembles a
narrow ring ($\Delta R$$\sim$35~AU) composed of large dust grains at the location
of the disk wall. Such narrow dust rings are also seen in other
transitional disks and may be evidence of dust trapping in pressure
bumps, possibly produced by planetary companions. More sensitive
submillimeter observations of the gas are necessary to further probe the
physical mechanisms at work in shaping the spatial distribution of large
dust in this disk. Interestingly, the IRAS~04125+2902 disk
is significantly fainter than other transitional disks that have been
resolved at submillimeter wavelengths, hinting that more
objects with large disk holes may exist at the faint end of
the submillimeter luminosity distribution that await detection with more
sensitive imaging telescopes.

\end{abstract}

\keywords{
planets and satellites: formation ---
protoplanetary disks  ---
planet-disk interactions  ---
stars: pre-main sequence  ---
infrared: planetary systems  ---
submillimeter: planetary systems
}

\section{Introduction} \label{intro}

Transitional disks have been singled out for their possible connection
to planet formation.  These disks have significant depletions of dust in
their inner regions and many researchers have proposed that such
clearing is due to planet-disk interactions \citep[see review by][]{espaillat14}. Potential protoplanets have also been identified in these
disks \citep[e.g.,][]{kraus11,huelamo11,reggiani14,biller14}, lending further support to the idea
that planets are leading to the observed disk clearings. However,
alternative mechanisms such as photoevaporation
\citep{hollenbach94,clarke01} can explain a subset of transitional disks
\citep{alexander09,owen11}.  A more detailed look at these objects is
warranted to understand the links to planet formation.

The detail in which we can study transitional disk structure has greatly
improved over time. The holes in transitional disks were first inferred
based on ``dips'' seen in the ground-based infrared (IR) spectral energy
distributions \citep[SEDs;][]{strom89,skrutskie90}. {\it Spitzer} later
greatly expanded the known number of transitional disks
\citep[e.g.,][]{muzerolle10,kim13} and provided much more detailed SEDs.
 {\it Spitzer} IRS allowed for detailed modeling
\citep[e.g.,][]{dalessio05}, as well as the detection of gaps in disks
(as opposed to holes) later dubbed ``pre-transitional''
\citep{espaillat07b}.  Such modeling inferred that the gaps and holes in
these pre-transitional disks and transitional disks were very large, of
the order of 10s of AU \citep{calvet05,brown07,espaillat07b}.
Submillimeter images confirmed these predictions by revealing large
cavities in the distribution of large grains in the disk
\citep[e.g.,][]{hughes07, brown08, hughes09, brown09, andrews09,
isella10b, andrews11}.

Resolved submillimeter interferometric imaging of
disks is greatly expanding our understanding of the spatial
distributions of dust and gas. Submillimeter
images have revealed that the large dust grains in the disk do not
extend as far out as the gas in some objects \citep[e.g.,][]{isella07,panic09, andrews12,
rosenfeld13,degregorio13,pineda14, pietu14}. The small dust disks seen in these observations
have been interpreted as evidence of dust evolution, in particular dust
radial drift \citep[e.g.,][]{weidenschilling77,barriere05,laibe08}. \citet{birnstiel14} showed that
dust radial drift will lead to the concentration of larger grains at
smaller radii, as well as a sharp outward truncation of the disk, as is
observed \citep[e.g.,][]{andrews12}.  This differentiates the signatures
of dust radial drift from those of dust grain growth, which should also
lead to the concentration of larger grains towards the inner radii the
disk, but not a sharp outward truncation of the disk
\citep{birnstiel14}. In the case of transitional disks, dust radial
drift can lead to concentrations of dust at the outer wall as the dust
gets trapped due to pressure bumps in the gas produced by planetary
companions \citep[e.g.,][]{zhu12,pinilla12}
and recent observations point to the detection of such ``dust traps'' 
in disks \citep[e.g.,][]{rosenfeld13,casassus13,
fukagawa13, vandermarel13, isella13, perez14, pineda14}.

Note that most of the above studies have focused on the brightest submillimeter
disks. Submillimeter observations to date have found that there is a
high frequency of large cavities amongst the more luminous half of the
disk population. \citet{andrews11} estimated that at least 1 in 5 of the
disks in the bright half of the Taurus and Ophiuchus submillimeter
luminosity distribution have cavities with radii $>$ 15~AU. This
suggests holes and gaps tend to be found in more massive disks, which
was also found in a demographic study of Taurus by \citet{najita07a}.
However, this may be due to a selection effect given that few faint
disks have been resolved.

Here we add observations of a faint \citep[$<$20~mJy at 1.3~mm;][]{andrews13} transitional disk in Taurus to
the growing literature on (pre-)transitional disks. We present an
SMA high-resolution 880 {\micron} image of IRAS~04125+2902 as
well as new {\it UVBRI} photometry from the DCT. 
IRAS~04125+2902 was first identified as a candidate member
of Taurus by \citet{kenyon94} and later confirmed as a member by
\citet{luhman09} who identified it as an M-type star. A transitional disk was
inferred using {\it Spitzer} IRAC and MIPS
photometry \citep{luhman09} and later {\it
Spitzer} IRS spectra \citep{furlan11}. IRAS~04125+2902 joins the previously imaged Taurus (pre-)transitional disks
around GM Aur, DM Tau, UX Tau A, LkCa 15, and RY Tau \citep{pietu06,
hughes09, isella10a, andrews11, andrews11b, isella12, isella14}.

This paper is organized into the following sections.  In Section~2, we
present the SMA and DCT observations of IRAS~04125+2902.
In Section~3, we constrain the stellar properties
and derive an accretion rate for IRAS~04125+2902.  We also report the results of simultaneous broad-band SED and
submillimeter visibility modeling. In Section~4, we discuss our results
in light of previous transitional disk studies.  Finally, in
Section~5, we summarize the key results of this work.

\section{Observations and Data Reduction} \label{redux}

\subsection{Submillimeter Data}

IRAS~04125+2902 was observed with the 8-element SMA 
interferometer on Mauna Kea, Hawaii using the 345 GHz dual-sideband receivers 
in the compact ($\sim$10--70 m baselines; 2011 Oct 26), extended ($\sim$30--230 
m baselines; 2011 Aug 19), and very extended ($\sim$70--500 m baselines; 2011 
Sep 8) configurations. The SMA correlator was configured to process 4 GHz of 
bandwidth per sideband, centered $\pm$6 GHz from the local oscillator frequency 
of 340.8 GHz (880 $\mu$m), divided into 48 spectral chunks containing 32 
coarse resolution (3.4 MHz) channels each. 

One chunk centered on the CO 
$J$=3$-$2 transition at 345.796 GHz had slightly higher resolution (corresponding to a 
velocity resolution of 0.7 km s$^{-1}$).  Note that we did not detect CO emission from our target.  We do not derive an upper limit since the line is not detected 
and there is substantial molecular cloud contamination in the CO line.
The observations cycled between IRAS~04125+2902 and the nearby quasars 3C 111 (1.8~Jy) and J0510+180 (0.9~Jy) over $\sim$10--15 
minute intervals. Bright quasars (3C 84, 3C 279) and primary flux calibration 
sources (Uranus, Callisto) were observed when the target had low elevation. 
Observing conditions were generally good, with precipitable water vapor levels 
under 2 mm. The raw visibilities were calibrated using the {\tt MIR}\footnote{http://www.cfa.harvard.edu/~cqi/mircook.html} software package. The passband 
response was corrected based on observations of bright quasars, and the 
amplitude scale was referenced to Uranus and Callisto. Complex gain calibration 
was performed using the repeated observations of the nearby quasars. The 
visibilities were spectrally binned into wideband continuum channels; the data 
from all three configurations were then combined. The calibrated visibilities 
were Fourier inverted (using a Briggs weighting with robust parameter of -1, chosen by experimentation to
represent the best combination of angular resolution and S$/$N for these data.), 
deconvolved, and then restored with a synthesized {\tt CLEAN} beam using 
standard tasks in the {\tt MIRIAD} package. The resulting image, shown in 
Figure 1, has a synthesized beam size of $0\farcs33\times0\farcs26$, 
and an RMS noise level of 1.3 mJy beam$^{-1}$.  We measure a flux density of
0.027$\pm$0.001~Jy for IRAS~04125+2902.
The flux density was estimated from the shortest visibility spacings, but
one finds the same values (within the uncertainties) from integrating the
synthesized images within the 1$\sigma$ contours or a reasonable aperture (1$\farcs$0 in
radius).

\begin{figure*}[ht!]
\epsscale{1.1}
\plotone{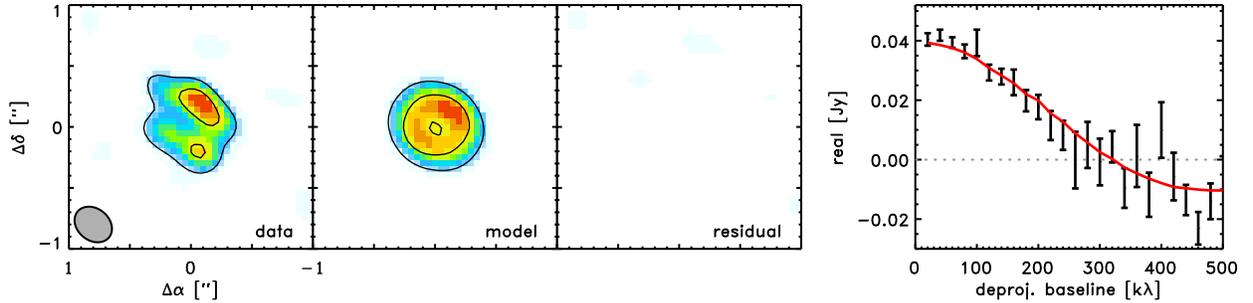}
\caption[]{SMA 880~{\micron} continuum emission of IRAS~04125+2902.
From the left, panels show the SMA continuum image, the synthetic model
image, the imaged residual, and the visibilities (black points)
with the model visibility (red line). Counter intervals are 3~${\sigma}$.
There is a double-peaked emission structure and a null in the
visibilities, indicating a large cavity in the disk.
}
\end{figure*}

\begin{deluxetable*}{lccccc}[ht!]
\tabletypesize{\scriptsize}
\tablewidth{0pt}
\tablecaption{Optical Photometry}
\startdata
\hline
\hline
Target & U  & B & V  &  R & I  \\
 \hline
IRAS~04125+2902 & $19.191\pm0.041$ & $17.442\pm0.011$ & $15.487\pm0.003$ & $13.954\pm0.005$ & $12.418\pm0.005$ 
\enddata
\end{deluxetable*}

\subsection{Optical Data}  

On 2013 Nov 29 we used the Large Monolithic Imager\footnote{http://www.lowell.edu/techSpecs/LMI/LMI.html} (LMI) on
the 4.3 m DCT to
obtain \emph{UBVRI} photometry of IRAS~04125+2902. The
FOV of the LMI is $12\farcm5$ $\times$ $12\farcm5$ ($0\farcs12$ per
unbinned pixel). We used 2 $\times$ 2 pixel binning for these
observations ($0\farcs24$ per pixel). The bias, flat-field, and overscan
calibration of the CCD images were performed using the \emph{ccdred}
package in IRAF. The photometric calibration of all images was carried
out interactively using the \emph{daophot} and \emph{photcal} packages
in IRAF, with standard stars selected from \citet{landolt92} and
following the procedure outlined in \citet{massey92}.  Photometry is
presented in Table~1 and Figure~2.  Note that our data of
IRAS~04125+2902 are consistent with previous $B$- (17.83, 17.44), $R$- (14.39, 14.34), 
and $I$-band (12.81) measurements 
from the US Naval Observatory catalog \citep{zacharias09} given the observational
and calibration uncertainties.   For clarity, we do not show these additional data in Figure~2.
Our IRAS~04125+2902 data are also consistent with previous $ugrz$ SDSS data \citep[Figure~2;][]{adelman11} 
within the uncertainties.

\section{Analysis \& Results}

In this section, we first review the disk model used in this work.  We
then discuss the adopted and derived stellar properties and accretion
rate, which are used as inputs in the model. Finally, we present the
first detailed modeling of IRAS~04125+2902. 
In short, we generated synthetic images for a grid of
disk models with the best chi-squared values from SED fitting and
compared them to the observed submm images to arrive at our best fit. 
Individually, SED modeling and submm image modeling have their
limitations, but below we discuss how modeling SEDs and submm images
together can help break some model degeneracies.

\subsection{Disk Model}

Here we use the disk models of
\citet{dalessio98,dalessio99,dalessio01,dalessio05,dalessio06}. 
The application of these models to the particular case of
fitting transitional disks has been discussed in detail before by
\citet{espaillat10}. In short, the D'Alessio et al. models are of
irradiated, accretion disks around pre-main-sequence stars.  The
temperature and density structure of the disk is calculated iteratively
and the surface density of the disk is tied to the accretion rate. A
transitional disk model has a hole relative to a full disk model
\citep{dalessio06}. There is a frontally illuminated ``wall'' at the
edge of the hole (i.e., at the inner edge of the disk). This
wall dominates the mid-infrared (MIR) emission (from $\sim$20--30~{\micron}) and the
disk dominates the emission beyond $\sim$40~{\micron}. 

The composition of dust used in the disk model impacts the resulting
emission and derived disk properties \citep[see][]{espaillat10}. Here we include silicates and graphite with
fractional abundances of 0.004 and 0.0025, respectively, following
the \citet{draine84} model for the diffuse ISM.  We calculate the silicate
(olivine) and graphite opacities using Mie theory and optical constants
from \citet{dorschner95} and \citet{draine84}, respectively. The models
assume spherical grains with a size distribution that scales like $a^{-p}$ between
grain radii of $a_{min}$ and $a_{max}$ and $p$ is 3.5
\citep{mathis77}. In the wall and the disk atmosphere,  $a_{min}$
is held fixed at 0.005~{\micron} while $a_{max}$ is varied 
to achieve the best fit to the SED,
particularly the 10~{\micron} and 20~{\micron} silicate emission
features. Near the disk midplane, the maximum grain size is held fixed at
1~mm \citep{dalessio06}.  

In fitting the outer wall of the disk, 
T$_{wall}$, the temperature at the surface of the optically thin
wall atmosphere, is adjusted to fit the SED.  
The height of the wall (z$_{wall}$) is equal to the disk scale height (H)
at the the radius in the disk at which the wall is located ($R_{wall}$). 
$R_{wall}$ is derived using
the best fitting $T_{wall}$ following
\begin{equation}R_{wall} \sim 
\left [{\frac{(L_* + L_{acc})}{16 \pi \sigma_R} }
( 2 + { \frac{\kappa_s} {\kappa_d} }) \right ] ^ {1/2} 
{1 \over T_{wall}^2 }.
\end{equation} 
where $\sigma_R$ is the Stefan-Boltzmann constant, $\tau_d$ is the total
mean optical depth in the disk, and $\kappa_s$ and $\kappa_d$ are the
mean opacities to the incident and local radiation, respectively. $L_* $
is the stellar luminosity and $L_{acc}$ ($\sim GM_*\mdot/R_*$) is the
luminosity of the stellar accretion shock. 

The parameters of the disk which are varied to fit the data are the
viscosity parameter ($\alpha$) and the settling parameter ($\epsilon$). 
One can interpret varying $\alpha$ for disks with the same inner and
outer radii as fitting for the disk mass since
M$_{disk}\propto{\mdot}/{\alpha}$ \citep[see Equation 38
in][]{dalessio98}. We note that the mass accretion rate onto the star
does not necessarily reflect the mass transport across the disk
nor do we expect that the mass accretion rate is constant throughout the
disk. However, for simplicity, here we assume that the mass accretion
rate measured onto the star is representative of the disk's accretion
rate. We parameterize settling as a depletion of dust in the upper
layers using $\epsilon=\zeta_{atm}/\zeta_{std}$ (i.e., the dust-to-gas
mass ratio of the disk atmosphere divided by the standard value), with a
corresponding increment of the dust-to-gas mass ratio near the midplane. 
One can also vary the disk inclination and outer radius ($R_{wall}$), which can
be constrained with high-resolution submillimeter images \citep[e.g.,][]{andrews11}.
Details of our fitting procedure in the particular case of
IRAS~04125+2902 can be found in Section~\ref{sec:datafitting}.

\subsection{Stellar Properties}

The star is a significant source of heating for the disk and so the
stellar properties are important input parameters for the code. The
stellar parameters used here are listed in Table 2. 
We use the same spectral type for IRAS~04125+2902
adopted in \citet{furlan11} and \citet{andrews13}, which originally comes from \citet{luhman09} using IR spectra and
with reported uncertainties of $\pm$0.25 subclasses. 
We measured a visual extinction by comparing observed colors to photospheric colors from
\citet{kh95}. We adopted an R$_V$ of 3.1 (A$_V$$/$A$_J$=3.55) and
dereddened our data using the extinction curve from \citet{mcclure09}
for IRAS~04125+2902 (A$_V$$<$ 3). Note
that our derived extinction value for IRAS~04125+2902 (A$_V$=2.7$\pm$0.5) agrees
with that found by \citet{furlan11} and \citet{andrews13}.  
The stellar temperature (T$_{*}$) is taken
from \citet{kh95}, based upon the adopted spectral type. 
The luminosity
is calculated with dereddened $J$-band photometry from 2MASS
\citep{skrutskie06} following \citet{kh95} assuming a distance of 140$\pm$20~pc
for Taurus \citep{bertout99}. 
R$_{*}$ is calculated using the derived
luminosity and adopted temperature. 
M$_{*}$ was derived with the
\citet{siess00} pre-main sequence evolutionary tracks using T$_{*}$ and
L$_{*}$.  
The uncertainties in L$_{*}$ and R$_{*}$ are dominated by the uncertainty of the extinction measurement.
We note that our values for the luminosity and mass are in good agreement with those
found by \citet{andrews13} with a Bayesian inference approach.

\begin{figure}
\epsscale{1.0}
\plotone{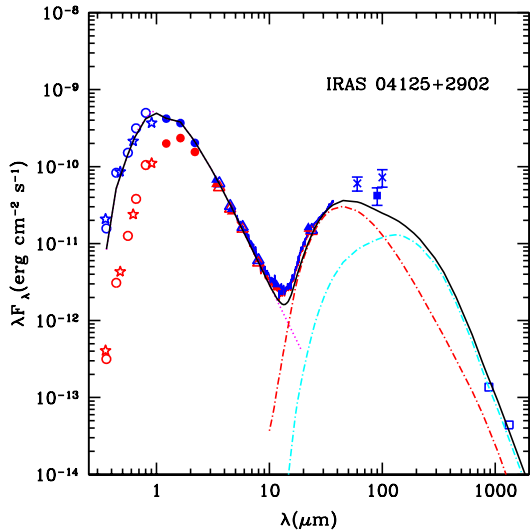}
\caption[]{Disk model (black solid line) fit to the dereddened SED (blue) of IRAS~04125+2902.
The original data (red)
were dereddened (blue) with the extinction listed in Table~1 and details
can be found in Section 3.2. Refer to Section~3.4 and Table~2 for model details. 
Here we show new DCT data (open circles; Table~1) from this work
and SMA (open squares) data first presented by \citet{andrews13}. 
We also include SDSS data \citep[open stars;][]{adelman11}, 
JHK data \citep[closed circles;][]{skrutskie06}, 
WISE \citep[closed triangles;][]{wright10},
IRAC and MIPS photometry \citep[open triangles;][]{luhman10,rebull10}, 
IRAS \citep[crosses][]{beichman88},
AKARI \citep[closed square;][]{ishihara10},
and IRS spectra \citep{furlan11}. 
The broken magenta line represents the stellar
photosphere based on colors from \citet{kh95} and scaled at $J$. 
Separate model components are
the stellar photosphere (magenta dotted line), the outer wall (red dot-short-dashed line), and the disk (cyan dot-dashed line). 
(A color version of this figure is
available in the online journal.)
}
\end{figure}

\begin{deluxetable}{lc}
\tabletypesize{\scriptsize}
\tablewidth{0pt}
\tablecaption{Stellar and Disk Properties of IRAS~04125+2902}
\startdata
\hline
\hline
\multicolumn{2}{c}{Stellar Properties}\\
\hline
M$_{*}$ (M$_{\sun}$)$^{1}$ & 0.5$\pm$0.1   \\ 
R$_{*}$ (R$_{\sun}$)$^{1}$ & 1.5$\pm$0.1   \\
T$_{*}$ (K)$^{2}$ & 3720$\pm$70  \\
L$_{*}$ ($\lsun$)$^{1}$ & 0.40$\pm$0.06   \\
A$_{V}$$^{1}$ & 2.7$\pm$0.5  \\
Spectral Type$^{2}$ & M1.25$\pm$0.25 \\
Inclination ($^{\circ}$)$^{3}$ & $<$30\\  
$\mdot$ (M$_{\sun}$ yr$^{-1}$)$^{1}$ & 3$_{-2}^{+6}$$\times$10$^{-10}$  \\
\hline
\hline
\multicolumn{2}{c}{Outer Wall \& Disk} \\
\hline
a$_{max}$ ({\micron})$^{3}$ & 10--100  \\ 
T$_{wall}$ (K)$^{3}$ & 85--95  \\ 
z$_{wall}$ (AU)$^{4}$ & 1.8--2.5 \\ 
R$_{wall}$ (AU)$^{4}$ & 18--24 \\ 
R$_{disk}$ (AU)$^{3}$ & 50--60 \\ 
$\epsilon$$^{3}$ & 0.01--1.0  \\ 
$\alpha$$^{3}$ & 0.0001--0.0003  \\ 
M$_{disk}$ (M$_{\sun}$)$^{4}$ & ~0.005--0.007  
\enddata
\tablenotetext{1}{These values were derived in this work; see Sections 3.2 and 3.3 for more details.
}
\tablenotetext{2}{The spectral type was adopted from \citet{luhman09} and the stellar temperature is from \citet{kh95} using this adopted spectral type.}
\tablenotetext{3}{These values are free parameters in the model and were derived based on fitting the SED and submm image; see Section 3.4 for more details.
}
\tablenotetext{4}{The wall height, wall radius, and disk mass were calculated based on the best-fitting disk parameters following the methods discussed in Section 3.1.
}
\end{deluxetable}

\subsection{Accretion Rate}

In the current paradigm of TTS accretion, material from the disk is
channeled onto the stellar surface via the stellar magnetospheric field
lines \citep[e.g.,][]{uchida84}, forming a hot accretion shock
\citep[$\sim$5,000-10,000 K][]{calvet98}.  Accretion rates can be
derived by measuring the accretion luminosity due to the accretion shock
on the stellar surface, which emits excess emission above the stellar
photosphere, that peaks in the ultraviolet \citep{calvet98}. $U$-band
excesses have been found to correlate with the total accretion shock
excess \citep{calvet98} and so $U$-band data can measure the ultraviolet
excess above the stellar photosphere
\citep[e.g.,][]{gullbring98,white01,rigliaco12,espaillat12}.

Here we measure an accretion rate for IRAS~04125+2902 of
3$\times$10$^{-10}$ M$_{\sun}$ yr$^{-1}$ using the DCT $U$-band data
presented in this work and the relation in \citet{gullbring98}.
This method has a typical uncertainty of a factor of 3 \citep{calvet04}.
This low accretion rate is consistent with IRAS~04125+2902's previously reported
H$\alpha$ equivalent width of 2.3$\pm$0.3 \citep{luhman09}, which is
expected from a non-accreting, weak-TTS \citep{white03}. Note that at
such low $\mdot$, $U$-band emission may be dominated by
chromospheric emission \citep{houdebine96,franchini98}, which can
contaminate mass accretion rate measurements \citep{ingleby11b}. Future
high-resolution ultraviolet spectra are needed to reliably constrain the
accretion rate of this object further.

\subsection{SED and Submillimeter Image Modeling} \label{sec:datafitting}

Using the stellar properties in Table~2, we ran a grid of $\sim$2,300 models where we
varied a$_{max}$ (5~{\micron}, 10~{\micron}, 100~{\micron}), T$_{wall}$ (75~K, 85~K, 95~K, 105~K), $\epsilon$ (0.01, 0.1, 0.5, 1.0), $\alpha$ (0.01, 0.001, 0.0001, 0.0003), R$_{disk}$ (40~AU, 50~AU, 60~AU, 100~AU), and the inclination (20$^{\circ}$, 40$^{\circ}$, 60$^{\circ}$).  
The parameters explored for a$_{max}$, T$_{wall}$, and R$_{disk}$ were constrained based on inspection of the SED and submm image.  The lack of strong silicate emission features in the IRS spectrum points to large ($>$5~{\micron}) grains in the disk atmosphere.  The steep slope of the IRS spectrum longwards of 10~{\microns} suggests cooler temperatures for the wall atmosphere ($>$100~K).  The submm image points to a relatively small ($<$100~AU) disk. 
The parameters explored for $\epsilon$, $\alpha$, and the inclination were chosen to cover a broad range of values with reasonable sampling while avoiding creating a computationally prohibitive grid.
We note that the above grid is not meant to be exhaustive or lead to a unique
fit.  We aim to identify the best fits to our data from
a reasonable parameter space meant to guide future observations.

For the models with the best chi-squared fits to the SED, high 
resolution model images at 880 $\mu$m were generated, and their Fourier 
transforms were sampled at the same spatial frequencies as the SMA 
observations. These model visibilities were then compared with the data in the 
form of an azimuthally-averaged deprojected visibilities 
\citep[see][]{andrews09}, as shown in the right-hand panel of Figure 1. Synthesis images were generated from the model and residual visibilities in 
the same way as for the data (middle panel of Figures 1).

A representative best fitting model to the submillimeter image and SED is
shown in Figures 1 -- 2, respectively. The particular model shown has R$_{wall}$=20~AU, R$_{disk}$=50~AU,
$\epsilon$ = 0.5, $\alpha$ = 0.0003, $i$=20$^{\circ}$, and a$_{max}$=100~{\micron}.  We do not show each of our best-fitting models since the fits are very similar.  Instead, in Table~2 we list the parameter ranges for the best fitting models.
The best fits to both the SED and submillimeter visibilities of IRAS~04125+2902
were achieved with disk models with wall radii of 18--24~AU and disk radii of 50--60~AU.  
We can also constrain the inclination of the disk of
IRAS~04125+2902 to $<$30$^{\circ}$. 
(We note that we arrived at this constraint by running additional models with inclinations between 0$^{\circ}$ and 40$^{\circ}$ in increments of 5$^{\circ}$ for the best fitting model shown in the figures.)
The wall and disk radii derived here as well as the inclinations were driven by the fit to the submillimeter visibilities since  
submm/mm imaging is sensitive to the spatial distribution of large dust in
the disk midplane while the SED is not as sensitive to the spatial distribution of dust.
The wall radius is determined from the null position in the visibilities. A11 showed that the submm/mm visibilities are sensitive to changes in the wall radius of about 10$\%$ given typical noise levels.
See A11 for a more in depth description regarding the implications of disk ring morphologies and nulls in submm visibilities.

The derived maximum grain size in the 
wall and disk atmosphere is largely based on fitting the 
IR SED since it is more sensitive to smaller dust grains
in the upper layers of the disk. 
In general, the dust grain size in the disk atmosphere can be constrained by the silicate emission features in the {\it Spitzer} IRS spectrum.  
Here we can get reasonable fits to the SED with a$_{max}$ between 10 to 100~{\micron}.
The $\alpha$ and $\epsilon$ parameters were constrained by fitting the IR and submm/mm SED data.
\citet{espaillat12} found that disk models with the same
$\epsilon$-to-$\alpha$ ratio will produce very similar emission in the IR but
substantially different emission in the submm$/$mm.  Hence, submm$/$mm data are necessary in conjunction with IR
data to break the degeneracy between dust settling and disk mass. Since in this work
we have both IR and submm$/$mm data, we can break this degeneracy. 
Here we find models with $\epsilon$ between 0.01 to 1 and $\alpha$ between 0.0001 and 0.0003 reproduce the SED and submm image best.  
Note that none of our models reproduce the IRAS/AKARI data
and we exclude this data from our chi-squared fitting.  It is unclear with the available data if the relatively high the IRAS/AKARI fluxes can be attributed to variability, as
has been seen in other (pre-)transitional disks \citep[e.g.,][]{muzerolle10,espaillat11,flaherty12}. 
We leave further exploration of this to future work.

We conclude that there is a relatively narrow
$\Delta R$$\sim$35~AU ring of large dust grains around IRAS~04125+2902.
Our submillimeter images are not of high enough
sensitivity to confirm or exclude disk asymmetries or to test
the extent of its dust disk relative to the gas disk. 
We leave this and more detailed studies
of the dust properties to future multi-wavelength work, which would
be ideally suited to probe this.

\section{Discussion} \label{sec:discuss}

IRAS~04125+2902 is one of the faintest transitional
disks imaged to date. \citet{andrews11} found that at least 1 in 5
submillimeter-bright, and therefore massive, disks are
(pre-)transitional disks. However, here we find a hole in a relatively
low submillimeter luminosity disk.  Given that the lower
luminosity end of the disk submillimeter brightness distribution has yet
to be probed, this suggests that many such disks may contain holes and gaps.

The dust in the disk around IRAS~04125+2902 only extends
from about 20~AU to 50--60~AU. This is reminiscent of a ring-like
large dust grain distribution. We stress that our SMA observations only trace
the submillimeter dust continuum and so this derived disk radius
applies only to the large dust grains in the disk and not necessarily the disk
overall.  Due to the faintness of this object, we do not have high
quality submillimeter data of the gas in the disk and cannot distinguish
between a disk where both the gas and dust extend over small spatial
scales or a disk where the dust is truncated to
significantly smaller radii than the gas.  With this in mind, below we
speculate on  the physical mechanisms that may underlie the observed
spatial distribution of dust in the disk of IRAS~04125+2902.  Higher sensitivity
observations of the gas, preferably targeting dense gas tracers (e.g., CO isotopologues) given
cloud contamination, 
are needed to explore the below scenarios further.

\subsection{Small disk radii}

Future submillimeter observations of the gas in the disk of
IRAS~04125+2902 may reveal that the gas traces the dust
distribution. In this case, such small gas and dust disks could be
attributed to a wide range of initial conditions at the time of disk
formation. \citet{andrews10} found an M$_{disk}$-R$_{disk}$ correlation
for disks in Ophiuchus in which fainter disks are smaller and less
massive. Even though this work uses a different disk model to derive disk
masses and outer radii, our results are roughly consistent with those of
\citet{andrews10}.  Those authors propose that this mass-radius
correlation could be the result of a wide range of initial conditions
and viscous properties at time of disk formation. 

Alternatively, smaller
disks could be the result of outward truncation by an unseen companion.
IRAS~04125+2902 is a
4$\farcs$0 binary, which corresponds to a separation of $\sim$560~AU at
140~pc (A. Kraus, 2014, private communication). This companion is unlikely to be responsible for
truncating the disk down to 50--60~AU according to the models of
\citet{artymowicz94} which show that the disk should be truncated to
0.2--0.5~$a$, where $a$ is the binary semi-major axis. 
However, a caveat is that we do not know the eccentricity of the companion's orbit
and disk truncation due to a highly eccentric companion ($e>$0.75) is not well constrained.
There are no equal brightness companions outside of $\sim$10~AU
and no companions near the substellar boundary outside $\sim$30~AU (A. Kraus, 2014, private communication).
Therefore, if the disk of IRAS~04125+2902 is being outwardly truncated by a companion, this companion
would have to be $<$50 M$_{Jupiter}$.

If the gas around IRAS~04125+2902
is found to extend to much larger radii than
the dust in the disk or multi-wavelength resolved data show a
size-wavelength anti-correlation, then the small dust disk observed
here may be evidence of dust evolution. Resolved submillimeter observations are increasingly
revealing that large dust grains in disks are distributed on more
compact spatial scales than the smaller grains and gas in the disk.  For
example, the mm-sized dust in the transitional disk of TW Hya is
truncated at 60 AU while CO gas emission is detected much further out in
the disk to about 215 AU \citep{andrews12}.  Evidence for compact dust
emission is also seen in full disks.  The full disk around AS 209 was
observed at four wavelengths between 0.88 mm and 9.8 mm; at longer
wavelengths, the dust disk is more compact, a signature that larger
grains are more concentrated towards the center of the disk
\citep{perez12}.

Observations revealing concentrations of larger dust grains towards the center of the disk
have been interpreted as evidence of dust evolution.   Larger dust grains
are not tied to the gas and are expected to drift radially inward
\citep{weidenschilling77}. This is due to the difference in velocities
between gas and dust of different sizes.  Dust travels at the Keplerian
velocity while the gas is partially pressure supported. In general, the
gas pressure decreases with radius so the gas pressure gradient is
negative and the gas travels at sub-Keplerian velocities.  Smaller dust
grains with slower radial velocities are coupled to the gas.  Larger
dust grains experience a headwind from the gas; the larger dust slows
down and loses angular momentum, spiraling in towards the star.  More
recent work by  \citet{birnstiel14} has shown that this process will not only lead to more
compact distributions for larger dust grains, but that there should also
be a distinct, sharp outer edge in the dust distribution, as is observed
\citep[e.g.,][]{andrews12}.  Grain growth, in contrast, would
lead to a more tapered edge \citep{birnstiel14}.

\subsection{Narrow dust rings}

The distribution of the large dust grains in IRAS~04125+2902 is narrow
(from about 20~AU to 50--60~AU) and ring-like.  As stated earlier, more sensitive
observations of the gas are needed to more firmly characterize the
disk's spatial properties. However, if it is revealed that the dust in
IRAS~04125+2902 is more compact than the gas, IRAS~04125+2902  would be
very similar to other transitional disks where concentrations of large
grains have been detected at the outer wall \citep[e.g.,][]{rosenfeld13}.
This has been interpreted as evidence of dust trappingâ due to pressure
bumps in the gas, possibly induced by a planet \citep{zhu12,pinilla12}.
In particular, simulations by \citet{pinilla12} show that this effect
can lead to narrow rings of $\sim$20~AU in width at the location of the outer
wall. In the transitional disk of V4046 Sgr, \citet{rosenfeld13} measure
a ring of 16~AU in width at 37~AU.  This is five times more compact than
the gas in this disk, suggesting an accumulation of solids at the local gas
maximum. In the pre-transitional disk of HD~100546, the CO (3-2) gas
extends out to 350 AU, while a $\sim$20~AU ring of large dust grains is
located at  $\sim$50 AU \citep{pineda14}. Similarly, narrow rings of large dust grains are
found in the
(pre-)transitional disks SAO 206462, SR 21 \citep{perez14}, and [PZ99]
J160421.7-213028 \citep{zhang14}.

The above scenario of a dust trap formed by planets is compatible with
the interpretation that the holes in disks are being carved out by
planets \citep[e.g.,][]{paardekooper04, zhu11, dodson11}. However, the
accretion rate we measure here for IRAS~04125+2902 is very low
(3$\times$10$^{-10}$ M$_{\sun}$ yr$^{-1}$), making the hole in this disk
a possible candidate for clearing due to photoevaporation
\citep[e.g.,][]{hollenbach94,clarke01,alexander06,alexander07,alexander14}. \citet{owen11}
found that they could explain disk holes with radii
$<$20~AU and accretion rates $<$10$^{-8}$ M$_{\sun}$ yr$^{-1}$ with
photoevaporative clearing. 
However, a low accretion rate is also compatible with clearing
due to a companion.
Besides CoKu Tau$/4$,  IRAS~04125+2902 has the lowest
measured accretion rate of the (pre-)transitional disks in Taurus
\citep[][Espaillat et al. 2014]{ingleby13}.  The low accretion rate  of CoKu Tau$/$4 
\citep[$<$10$^{-10}$
M$_{\sun}$ yr$^{-1}$;][]{cohen79} and hole size \citep[$\sim$14~AU;][]{nagel10} are
consistent with clearing due to the stellar mass companion detected
within the disk hole
\citep{ireland08,artymowicz94,lubow99}. 
More work is needed to exclude or detect companions within the disk
hole of IRAS~04125+2902.

Based on the above, IRAS~04125+2902's
accretion rate may indicate its hole is due to disk dispersal by
photoevaporation or there is a massive companion in the inner disk. The
first interpretation would be consistent with the proposal of
\citet{owen12b} that there are two distinct populations of transitional
disks - those with low mm fluxes,  small $<$10 AU holes, and low
($<$10$^{-9}$ M$_{\sun}$ yr$^{-1}$) accretion rates and those with high mm
fluxes, larger holes, and higher accretion rates.  The former population
is consistent with models of photoevaporation while the latter are not
\citep{owen11,owen12b}.  The former are also consistent with holes in disks
developing later in the lifetime of the disk.  Interestingly, Taurus
has both populations. The second interpretation of
dynamical clearing by a stellar or planetary mass companion
would need gas observations to confirm that
indeed a narrow ring of dust and gas exists at larger radii since, to the best
of our knowledge, photoevaporation models do not predict
such a feature.

Note that along with narrow rings in (pre-)transitional disks,
asymmetries have also been detected, which have been
interpreted as evidence of planet-induced vortices \citep{perez14, casassus13,
fukagawa13, vandermarel13, isella13, pineda14}. 
Our observations are not of high enough signal-to-noise to confirm or
exclude the presence of asymmetries in IRAS~04125+2902.

\section{Summary} \label{sec:summ}

We performed the first detailed characterization of the recently
identified transitional disk around IRAS 04125+2902 in
Taurus. We presented a new SMA high-resolution
880~{\micron} image and DCT photometry.  Using simultaneous SED
and submillimeter visibilities modeling, we find the following:

1. IRAS 04125+2902 is the faintest transitional
disk imaged in detail to date, suggesting that there are more faint, less massive
disks with inner holes that have not yet been detected.

2. IRAS 04125+2902 has a dust disk radius of 50--60 AU. Gas observations of IRAS~04125+2902 are necessary to explore the spatial distribution of the dust
relative to the gas.  If the dust and gas are both truncated to
similarly small outer radii, this could point to a wide range of initial
conditions in disks or an unseen companion.  If the dust is significantly more compact than the
gas in the disk, this could be evidence of dust radial drift in disks.

3. IRAS 04125+2902's dust is distributed as a narrow ring ($\Delta
R$$\sim$35~AU) of large grains at the location of the disk
wall ($\sim$20~AU). Such narrow dust rings are also seen in other transitional disks
and may be evidence of dust trapping in pressure bumps, possibly
produced by planetary companions.  However, here we also measure a very
low accretion rate for IRAS~04125+2902 of $\sim$3$\times$10$^{-10}$ M$_{\sun}$
yr$^{-1}$, making this object a potential candidate for clearing due to
photoevaporation.  More sensitive submillimeter observations of the gas
in this disk should be able to distinguish between clearing due to
planets or photoevaporation in the disk.

ALMA will play an important role in probing the lower luminosity end of
the transitional disk population and revealing the true frequency of
small dust disks and narrow rings in transitional disks.

 \acknowledgments{
 We thank the referee for comments that helped to improve the paper.
 We thank Nuria Calvet, Ramiro Franco-Hern{\`a}ndez, Elise Furlan, and Melissa McClure for discussions.  We thank Adam Kraus for sharing results
 before publication. The SMA is a joint project between the Smithsonian Astrophysical
 Observatory and the Academia Sinica Institute of Astronomy and
 Astrophysics and is funded by the Smithsonian Institution and the
 Academia Sinica. The authors wish to recognize and acknowledge the very significant cultural role and reverence that the summit of Mauna Kea has always had within the indigenous Hawaiian community.  We are most fortunate to have the opportunity to conduct observations from this mountain.  These results made use of the Discovery Channel
 Telescope at Lowell Observatory, supported by Discovery Communications,
 Inc., Boston University, the University of Maryland, the University of
 Toledo and Northern Arizona University. Finally, a special recognition
 of the contribution of Paola D'Alessio, who passed away in November of
 2013. She is greatly missed as a scientist, colleague, and friend. 
 Paola would have been happy to see her research continue to have an
 important impact in the star formation community.
}

{\it Facilities:} \facility{DCT (LMI)}, \facility{SMA}



\bibliographystyle{apjv2}

\begin{thebibliography}{103}
\expandafter\ifx\csname natexlab\endcsname\relax\def\natexlab#1{#1}\fi

\bibitem[{{Adelman-McCarthy} {et al.}(2011)}]{adelman11}
{Adelman-McCarthy}, J.~K., \& {et al.} 2011, VizieR Online Data Catalog, 2306,
  0

\bibitem[{{Alexander} {et~al.}(2013){Alexander}, {Pascucci}, {Andrews},
  {Armitage}, \& {Cieza}}]{alexander14}
{Alexander}, R., {Pascucci}, I., {Andrews}, S., {Armitage}, P., \& {Cieza}, L.
  2013, ArXiv e-prints

\bibitem[{{Alexander} \& {Armitage}(2007)}]{alexander07}
{Alexander}, R.~D., \& {Armitage}, P.~J. 2007, \mnras, 375, 500

\bibitem[{{Alexander} \& {Armitage}(2009)}]{alexander09}
---. 2009, \apj, 704, 989

\bibitem[{{Alexander} {et~al.}(2006){Alexander}, {Clarke}, \&
  {Pringle}}]{alexander06}
{Alexander}, R.~D., {Clarke}, C.~J., \& {Pringle}, J.~E. 2006, \mnras, 369, 229

\bibitem[{{Andrews} {et~al.}(2013){Andrews}, {Rosenfeld}, {Kraus}, \&
  {Wilner}}]{andrews13}
{Andrews}, S.~M., {Rosenfeld}, K.~A., {Kraus}, A.~L., \& {Wilner}, D.~J. 2013,
  \apj, 771, 129

\bibitem[{{Andrews} {et~al.}(2011{\natexlab{a}}){Andrews}, {Rosenfeld},
  {Wilner}, \& {Bremer}}]{andrews11b}
{Andrews}, S.~M., {Rosenfeld}, K.~A., {Wilner}, D.~J., \& {Bremer}, M.
  2011{\natexlab{a}}, \apjl, 742, L5

\bibitem[{{Andrews} {et~al.}(2011{\natexlab{b}}){Andrews}, {Wilner},
  {Espaillat}, {Hughes}, {Dullemond}, {McClure}, {Qi}, \& {Brown}}]{andrews11}
{Andrews}, S.~M., {Wilner}, D.~J., {Espaillat}, C., {Hughes}, A.~M.,
  {Dullemond}, C.~P., {McClure}, M.~K., {Qi}, C., \& {Brown}, J.~M.
  2011{\natexlab{b}}, \apj, 732, 42

\bibitem[{{Andrews} {et~al.}(2009){Andrews}, {Wilner}, {Hughes}, {Qi}, \&
  {Dullemond}}]{andrews09}
{Andrews}, S.~M., {Wilner}, D.~J., {Hughes}, A.~M., {Qi}, C., \& {Dullemond},
  C.~P. 2009, \apj, 700, 1502

\bibitem[{{Andrews} {et~al.}(2010){Andrews}, {Wilner}, {Hughes}, {Qi}, \&
  {Dullemond}}]{andrews10}
---. 2010, \apj, 723, 1241

\bibitem[{{Andrews} {et~al.}(2012){Andrews}, {Wilner}, {Hughes}, {Qi},
  {Rosenfeld}, {{\"O}berg}, {Birnstiel}, {Espaillat}, {Cieza}, {Williams},
  {Lin}, \& {Ho}}]{andrews12}
{Andrews}, S.~M., {et~al.} 2012, \apj, 744, 162

\bibitem[{{Artymowicz} \& {Lubow}(1994)}]{artymowicz94}
{Artymowicz}, P., \& {Lubow}, S.~H. 1994, \apj, 421, 651

\bibitem[{{Barri{\`e}re-Fouchet} {et~al.}(2005){Barri{\`e}re-Fouchet},
  {Gonzalez}, {Murray}, {Humble}, \& {Maddison}}]{barriere05}
{Barri{\`e}re-Fouchet}, L., {Gonzalez}, J.-F., {Murray}, J.~R., {Humble},
  R.~J., \& {Maddison}, S.~T. 2005, \aap, 443, 185

\bibitem[{{Beichman} {et~al.}(1988){Beichman}, {Neugebauer}, {Habing}, {Clegg},
  \& {Chester}}]{beichman88}
{Beichman}, C.~A., {Neugebauer}, G., {Habing}, H.~J., {Clegg}, P.~E., \&
  {Chester}, T.~J., eds. 1988, {Infrared astronomical satellite (IRAS) catalogs
  and atlases. Volume 1: Explanatory supplement}, Vol.~1

\bibitem[{{Bertout} {et~al.}(1999){Bertout}, {Robichon}, \&
  {Arenou}}]{bertout99}
{Bertout}, C., {Robichon}, N., \& {Arenou}, F. 1999, \aap, 352, 574

\bibitem[{{Biller} {et~al.}(2014){Biller}, {Males}, {Rodigas}, {Morzinski},
  {Close}, {Juh{\'a}sz}, {Follette}, {Lacour}, {Benisty}, {Sicilia-Aguilar},
  {Hinz}, {Weinberger}, {Henning}, {Pott}, {Bonnefoy}, \&
  {K{\"o}hler}}]{biller14}
{Biller}, B.~A., {et~al.} 2014, \apjl, 792, L22

\bibitem[{{Birnstiel} \& {Andrews}(2014)}]{birnstiel14}
{Birnstiel}, T., \& {Andrews}, S.~M. 2014, \apj, 780, 153

\bibitem[{{Brown} {et~al.}(2008){Brown}, {Blake}, {Qi}, {Dullemond}, \&
  {Wilner}}]{brown08}
{Brown}, J.~M., {Blake}, G.~A., {Qi}, C., {Dullemond}, C.~P., \& {Wilner},
  D.~J. 2008, \apjl, 675, L109

\bibitem[{{Brown} {et~al.}(2009){Brown}, {Blake}, {Qi}, {Dullemond}, {Wilner},
  \& {Williams}}]{brown09}
{Brown}, J.~M., {Blake}, G.~A., {Qi}, C., {Dullemond}, C.~P., {Wilner}, D.~J.,
  \& {Williams}, J.~P. 2009, \apj, 704, 496

\bibitem[{{Brown} {et~al.}(2007){Brown}, {Blake}, {Dullemond}, {Mer{\'{\i}}n},
  {Augereau}, {Boogert}, {Evans}, {Geers}, {Lahuis}, {Kessler-Silacci},
  {Pontoppidan}, \& {van Dishoeck}}]{brown07}
{Brown}, J.~M., {et~al.} 2007, \apjl, 664, L107

\bibitem[{{Calvet} \& {Gullbring}(1998)}]{calvet98}
{Calvet}, N., \& {Gullbring}, E. 1998, \apj, 509, 802

\bibitem[{{Calvet} {et~al.}(2004){Calvet}, {Muzerolle}, {Brice{\~n}o},
  {Hern{\'a}ndez}, {Hartmann}, {Saucedo}, \& {Gordon}}]{calvet04}
{Calvet}, N., {Muzerolle}, J., {Brice{\~n}o}, C., {Hern{\'a}ndez}, J.,
  {Hartmann}, L., {Saucedo}, J.~L., \& {Gordon}, K.~D. 2004, \aj, 128, 1294

\bibitem[{{Calvet} {et~al.}(2005){Calvet}, {D'Alessio}, {Watson},
  {Franco-Hern{\'a}ndez}, {Furlan}, {Green}, {Sutter}, {Forrest}, {Hartmann},
  {Uchida}, {Keller}, {Sargent}, {Najita}, {Herter}, {Barry}, \&
  {Hall}}]{calvet05}
{Calvet}, N., {et~al.} 2005, \apjl, 630, L185

\bibitem[{{Casassus} {et~al.}(2013){Casassus}, {van der Plas}, {M}, {Dent},
  {Fomalont}, {Hagelberg}, {Hales}, {Jord{\'a}n}, {Mawet}, {M{\'e}nard},
  {Wootten}, {Wilner}, {Hughes}, {Schreiber}, {Girard}, {Ercolano}, {Canovas},
  {Rom{\'a}n}, \& {Salinas}}]{casassus13}
{Casassus}, S., {et~al.} 2013, \nat, 493, 191

\bibitem[{{Clarke} {et~al.}(2001){Clarke}, {Gendrin}, \&
  {Sotomayor}}]{clarke01}
{Clarke}, C.~J., {Gendrin}, A., \& {Sotomayor}, M. 2001, \mnras, 328, 485

\bibitem[{{Cohen} \& {Kuhi}(1979)}]{cohen79}
{Cohen}, M., \& {Kuhi}, L.~V. 1979, \apjs, 41, 743

\bibitem[{{D'Alessio} {et~al.}(2001){D'Alessio}, {Calvet}, \&
  {Hartmann}}]{dalessio01}
{D'Alessio}, P., {Calvet}, N., \& {Hartmann}, L. 2001, \apj, 553, 321

\bibitem[{{D'Alessio} {et~al.}(2006){D'Alessio}, {Calvet}, {Hartmann},
  {Franco-Hern{\'a}ndez}, \& {Serv{\'{\i}}n}}]{dalessio06}
{D'Alessio}, P., {Calvet}, N., {Hartmann}, L., {Franco-Hern{\'a}ndez}, R., \&
  {Serv{\'{\i}}n}, H. 2006, \apj, 638, 314

\bibitem[{{D'Alessio} {et~al.}(1999){D'Alessio}, {Calvet}, {Hartmann},
  {Lizano}, \& {Cant{\'o}}}]{dalessio99}
{D'Alessio}, P., {Calvet}, N., {Hartmann}, L., {Lizano}, S., \& {Cant{\'o}}, J.
  1999, \apj, 527, 893

\bibitem[{{D'Alessio} {et~al.}(1998){D'Alessio}, {Canto}, {Calvet}, \&
  {Lizano}}]{dalessio98}
{D'Alessio}, P., {Canto}, J., {Calvet}, N., \& {Lizano}, S. 1998, \apj, 500,
  411

\bibitem[{{D'Alessio} {et~al.}(2005){D'Alessio}, {Hartmann}, {Calvet},
  {Franco-Hern{\'a}ndez}, {Forrest}, {Sargent}, {Furlan}, {Uchida}, {Green},
  {Watson}, {Chen}, {Kemper}, {Sloan}, \& {Najita}}]{dalessio05}
{D'Alessio}, P., {et~al.} 2005, \apj, 621, 461

\bibitem[{{de Gregorio-Monsalvo} {et~al.}(2013){de Gregorio-Monsalvo},
  {M{\'e}nard}, {Dent}, {Pinte}, {L{\'o}pez}, {Klaassen}, {Hales},
  {Cort{\'e}s}, {Rawlings}, {Tachihara}, {Testi}, {Takahashi}, {Chapillon},
  {Mathews}, {Juhasz}, {Akiyama}, {Higuchi}, {Saito}, {Nyman}, {Phillips},
  {Rod{\'o}n}, {Corder}, \& {Van Kempen}}]{degregorio13}
{de Gregorio-Monsalvo}, I., {et~al.} 2013, \aap, 557, A133

\bibitem[{{Dodson-Robinson} \& {Salyk}(2011)}]{dodson11}
{Dodson-Robinson}, S.~E., \& {Salyk}, C. 2011, \apj, 738, 131

\bibitem[{{Dorschner} {et~al.}(1995){Dorschner}, {Begemann}, {Henning},
  {Jaeger}, \& {Mutschke}}]{dorschner95}
{Dorschner}, J., {Begemann}, B., {Henning}, T., {Jaeger}, C., \& {Mutschke}, H.
  1995, \aap, 300, 503

\bibitem[{{Draine} \& {Lee}(1984)}]{draine84}
{Draine}, B.~T., \& {Lee}, H.~M. 1984, \apj, 285, 89

\bibitem[{{Espaillat} {et~al.}(2007){Espaillat}, {Calvet}, {D'Alessio},
  {Hern{\'a}ndez}, {Qi}, {Hartmann}, {Furlan}, \& {Watson}}]{espaillat07b}
{Espaillat}, C., {Calvet}, N., {D'Alessio}, P., {Hern{\'a}ndez}, J., {Qi}, C.,
  {Hartmann}, L., {Furlan}, E., \& {Watson}, D.~M. 2007, \apjl, 670, L135

\bibitem[{{Espaillat} {et~al.}(2011){Espaillat}, {Furlan}, {D'Alessio},
  {Sargent}, {Nagel}, {Calvet}, {Watson}, \& {Muzerolle}}]{espaillat11}
{Espaillat}, C., {Furlan}, E., {D'Alessio}, P., {Sargent}, B., {Nagel}, E.,
  {Calvet}, N., {Watson}, D.~M., \& {Muzerolle}, J. 2011, \apj, 728, 49

\bibitem[{{Espaillat} {et~al.}(2010){Espaillat}, {D'Alessio}, {Hern{\'a}ndez},
  {Nagel}, {Luhman}, {Watson}, {Calvet}, {Muzerolle}, \&
  {McClure}}]{espaillat10}
{Espaillat}, C., {et~al.} 2010, \apj, 717, 441

\bibitem[{{Espaillat} {et~al.}(2012){Espaillat}, {Ingleby}, {Hern{\'a}ndez},
  {Furlan}, {D'Alessio}, {Calvet}, {Andrews}, {Muzerolle}, {Qi}, \&
  {Wilner}}]{espaillat12}
---. 2012, \apj, 747, 103

\bibitem[{{Espaillat} {et~al.}(2014){Espaillat}, {Muzerolle}, {Najita},
  {Andrews}, {Zhu}, {Calvet}, {Kraus}, {Hashimoto}, {Kraus}, \&
  {D'Alessio}}]{espaillat14}
---. 2014, ArXiv e-prints

\bibitem[{{Flaherty} {et~al.}(2012){Flaherty}, {Muzerolle}, {Rieke},
  {Gutermuth}, {Balog}, {Herbst}, {Megeath}, \& {Kun}}]{flaherty12}
{Flaherty}, K.~M., {Muzerolle}, J., {Rieke}, G., {Gutermuth}, R., {Balog}, Z.,
  {Herbst}, W., {Megeath}, S.~T., \& {Kun}, M. 2012, \apj, 748, 71

\bibitem[{{Franchini} {et~al.}(1998){Franchini}, {Morossi}, \&
  {Malagnini}}]{franchini98}
{Franchini}, M., {Morossi}, C., \& {Malagnini}, M.~L. 1998, \apj, 508, 370

\bibitem[{{Fukagawa} {et~al.}(2013){Fukagawa}, {Tsukagoshi}, {Momose}, {Saigo},
  {Ohashi}, {Kitamura}, {Inutsuka}, {Muto}, {Nomura}, {Takeuchi}, {Kobayashi},
  {Hanawa}, {Akiyama}, {Honda}, {Fujiwara}, {Kataoka}, {Takahashi}, \&
  {Shibai}}]{fukagawa13}
{Fukagawa}, M., {et~al.} 2013, \pasj, 65, L14

\bibitem[{{Furlan} {et~al.}(2011){Furlan}, {Luhman}, {Espaillat}, {D'Alessio},
  {Adame}, {Manoj}, {Kim}, {Watson}, {Forrest}, {McClure}, {Calvet}, {Sargent},
  {Green}, \& {Fischer}}]{furlan11}
{Furlan}, E., {et~al.} 2011, \apjs, 195, 3

\bibitem[{{Gullbring} {et~al.}(1998){Gullbring}, {Hartmann}, {Briceno}, \&
  {Calvet}}]{gullbring98}
{Gullbring}, E., {Hartmann}, L., {Briceno}, C., \& {Calvet}, N. 1998, \apj,
  492, 323

\bibitem[{{Hollenbach} {et~al.}(1994){Hollenbach}, {Johnstone}, {Lizano}, \&
  {Shu}}]{hollenbach94}
{Hollenbach}, D., {Johnstone}, D., {Lizano}, S., \& {Shu}, F. 1994, \apj, 428,
  654

\bibitem[{{Houdebine} {et~al.}(1996){Houdebine}, {Mathioudakis}, {Doyle}, \&
  {Foing}}]{houdebine96}
{Houdebine}, E.~R., {Mathioudakis}, M., {Doyle}, J.~G., \& {Foing}, B.~H. 1996,
  \aap, 305, 209

\bibitem[{{Hu{\'e}lamo} {et~al.}(2011){Hu{\'e}lamo}, {Lacour}, {Tuthill},
  {Ireland}, {Kraus}, \& {Chauvin}}]{huelamo11}
{Hu{\'e}lamo}, N., {Lacour}, S., {Tuthill}, P., {Ireland}, M., {Kraus}, A., \&
  {Chauvin}, G. 2011, \aap, 528, L7

\bibitem[{{Hughes} {et~al.}(2007){Hughes}, {Wilner}, {Calvet}, {D'Alessio},
  {Claussen}, \& {Hogerheijde}}]{hughes07}
{Hughes}, A.~M., {Wilner}, D.~J., {Calvet}, N., {D'Alessio}, P., {Claussen},
  M.~J., \& {Hogerheijde}, M.~R. 2007, \apj, 664, 536

\bibitem[{{Hughes} {et~al.}(2009){Hughes}, {Andrews}, {Espaillat}, {Wilner},
  {Calvet}, {D'Alessio}, {Qi}, {Williams}, \& {Hogerheijde}}]{hughes09}
{Hughes}, A.~M., {et~al.} 2009, \apj, 698, 131

\bibitem[{{Ingleby} {et~al.}(2011){Ingleby}, {Calvet}, {Bergin}, {Herczeg},
  {Brown}, {Alexander}, {Edwards}, {Espaillat}, {France}, {Gregory},
  {Hillenbrand}, {Roueff}, {Valenti}, {Walter}, {Johns-Krull}, {Brown},
  {Linsky}, {McClure}, {Ardila}, {Abgrall}, {Bethell}, {Hussain}, \&
  {Yang}}]{ingleby11b}
{Ingleby}, L., {et~al.} 2011, \apj, 743, 105

\bibitem[{{Ingleby} {et~al.}(2013){Ingleby}, {Calvet}, {Herczeg}, {Blaty},
  {Walter}, {Ardila}, {Alexander}, {Edwards}, {Espaillat}, {Gregory},
  {Hillenbrand}, \& {Brown}}]{ingleby13}
---. 2013, \apj, 767, 112

\bibitem[{{Ireland} \& {Kraus}(2008)}]{ireland08}
{Ireland}, M.~J., \& {Kraus}, A.~L. 2008, \apjl, 678, L59

\bibitem[{{Isella} {et~al.}(2010{\natexlab{a}}){Isella}, {Carpenter}, \&
  {Sargent}}]{isella10a}
{Isella}, A., {Carpenter}, J.~M., \& {Sargent}, A.~I. 2010{\natexlab{a}}, \apj,
  714, 1746

\bibitem[{{Isella} {et~al.}(2014){Isella}, {Chandler}, {Carpenter},
  {P{\'e}rez}, \& {Ricci}}]{isella14}
{Isella}, A., {Chandler}, C.~J., {Carpenter}, J.~M., {P{\'e}rez}, L.~M., \&
  {Ricci}, L. 2014, \apj, 788, 129

\bibitem[{{Isella} {et~al.}(2010{\natexlab{b}}){Isella}, {Natta}, {Wilner},
  {Carpenter}, \& {Testi}}]{isella10b}
{Isella}, A., {Natta}, A., {Wilner}, D., {Carpenter}, J.~M., \& {Testi}, L.
  2010{\natexlab{b}}, \apj, 725, 1735

\bibitem[{{Isella} {et~al.}(2012){Isella}, {P{\'e}rez}, \&
  {Carpenter}}]{isella12}
{Isella}, A., {P{\'e}rez}, L.~M., \& {Carpenter}, J.~M. 2012, \apj, 747, 136

\bibitem[{{Isella} {et~al.}(2013){Isella}, {P{\'e}rez}, {Carpenter}, {Ricci},
  {Andrews}, \& {Rosenfeld}}]{isella13}
{Isella}, A., {P{\'e}rez}, L.~M., {Carpenter}, J.~M., {Ricci}, L., {Andrews},
  S., \& {Rosenfeld}, K. 2013, \apj, 775, 30

\bibitem[{{Isella} {et~al.}(2007){Isella}, {Testi}, {Natta}, {Neri}, {Wilner},
  \& {Qi}}]{isella07}
{Isella}, A., {Testi}, L., {Natta}, A., {Neri}, R., {Wilner}, D., \& {Qi}, C.
  2007, \aap, 469, 213

\bibitem[{{Ishihara} {et~al.}(2010){Ishihara}, {Onaka}, {Kataza}, {Salama},
  {Alfageme}, {Cassatella}, {Cox}, {Garc{\'{\i}}a-Lario}, {Stephenson},
  {Cohen}, {Fujishiro}, {Fujiwara}, {Hasegawa}, {Ita}, {Kim}, {Matsuhara},
  {Murakami}, {M{\"u}ller}, {Nakagawa}, {Ohyama}, {Oyabu}, {Pyo}, {Sakon},
  {Shibai}, {Takita}, {Tanab{\'e}}, {Uemizu}, {Ueno}, {Usui}, {Wada},
  {Watarai}, {Yamamura}, \& {Yamauchi}}]{ishihara10}
{Ishihara}, D., {et~al.} 2010, \aap, 514, A1

\bibitem[{{Kenyon} {et~al.}(1994){Kenyon}, {Dobrzycka}, \&
  {Hartmann}}]{kenyon94}
{Kenyon}, S.~J., {Dobrzycka}, D., \& {Hartmann}, L. 1994, \aj, 108, 1872

\bibitem[{{Kenyon} \& {Hartmann}(1995)}]{kh95}
{Kenyon}, S.~J., \& {Hartmann}, L. 1995, \apjs, 101, 117

\bibitem[{{Kim} {et~al.}(2013){Kim}, {Watson}, {Manoj}, {Forrest}, {Najita},
  {Furlan}, {Sargent}, {Espaillat}, {Muzerolle}, {Megeath}, {Calvet}, {Green},
  \& {Arnold}}]{kim13}
{Kim}, K.~H., {et~al.} 2013, \apj, 769, 149

\bibitem[{{Kraus} {et~al.}(2011){Kraus}, {Ireland}, {Martinache}, \&
  {Hillenbrand}}]{kraus11}
{Kraus}, A.~L., {Ireland}, M.~J., {Martinache}, F., \& {Hillenbrand}, L.~A.
  2011, \apj, 731, 8

\bibitem[{{Laibe} {et~al.}(2008){Laibe}, {Gonzalez}, {Fouchet}, \&
  {Maddison}}]{laibe08}
{Laibe}, G., {Gonzalez}, J.-F., {Fouchet}, L., \& {Maddison}, S.~T. 2008, \aap,
  487, 265

\bibitem[{{Landolt}(1992)}]{landolt92}
{Landolt}, A.~U. 1992, \aj, 104, 340

\bibitem[{{Lubow} {et~al.}(1999){Lubow}, {Seibert}, \& {Artymowicz}}]{lubow99}
{Lubow}, S.~H., {Seibert}, M., \& {Artymowicz}, P. 1999, \apj, 526, 1001

\bibitem[{{Luhman} {et~al.}(2010){Luhman}, {Allen}, {Espaillat}, {Hartmann}, \&
  {Calvet}}]{luhman10}
{Luhman}, K.~L., {Allen}, P.~R., {Espaillat}, C., {Hartmann}, L., \& {Calvet},
  N. 2010, \apjs, 186, 111

\bibitem[{{Luhman} {et~al.}(2009){Luhman}, {Mamajek}, {Allen}, \&
  {Cruz}}]{luhman09}
{Luhman}, K.~L., {Mamajek}, E.~E., {Allen}, P.~R., \& {Cruz}, K.~L. 2009, \apj,
  703, 399

\bibitem[{{Massey} \& {Davis}(1992)}]{massey92}
{Massey}, P., \& {Davis}, L.~E. 1992, A UserÕs Guide to Stellar CCD Photometry
  with IRAF, http://iraf.noao.edu/iraf/ftp/iraf/docs/daophot2.ps.Z

\bibitem[{{Mathis} {et~al.}(1977){Mathis}, {Rumpl}, \& {Nordsieck}}]{mathis77}
{Mathis}, J.~S., {Rumpl}, W., \& {Nordsieck}, K.~H. 1977, \apj, 217, 425

\bibitem[{{McClure}(2009)}]{mcclure09}
{McClure}, M. 2009, \apjl, 693, L81

\bibitem[{{Muzerolle} {et~al.}(2010){Muzerolle}, {Allen}, {Megeath},
  {Hern{\'a}ndez}, \& {Gutermuth}}]{muzerolle10}
{Muzerolle}, J., {Allen}, L.~E., {Megeath}, S.~T., {Hern{\'a}ndez}, J., \&
  {Gutermuth}, R.~A. 2010, \apj, 708, 1107

\bibitem[{{Nagel} {et~al.}(2010){Nagel}, {D'Alessio}, {Calvet}, {Espaillat},
  {Sargent}, {Hern{\'a}ndez}, \& {Forrest}}]{nagel10}
{Nagel}, E., {D'Alessio}, P., {Calvet}, N., {Espaillat}, C., {Sargent}, B.,
  {Hern{\'a}ndez}, J., \& {Forrest}, W.~J. 2010, \apj, 708, 38

\bibitem[{{Najita} {et~al.}(2007){Najita}, {Strom}, \& {Muzerolle}}]{najita07a}
{Najita}, J.~R., {Strom}, S.~E., \& {Muzerolle}, J. 2007, \mnras, 378, 369

\bibitem[{{Owen} \& {Clarke}(2012)}]{owen12b}
{Owen}, J.~E., \& {Clarke}, C.~J. 2012, \mnras, 426, L96

\bibitem[{{Owen} {et~al.}(2011){Owen}, {Ercolano}, \& {Clarke}}]{owen11}
{Owen}, J.~E., {Ercolano}, B., \& {Clarke}, C.~J. 2011, \mnras, 412, 13

\bibitem[{{Paardekooper} \& {Mellema}(2004)}]{paardekooper04}
{Paardekooper}, S.-J., \& {Mellema}, G. 2004, \aap, 425, L9

\bibitem[{{Pani{\'c}} {et~al.}(2009){Pani{\'c}}, {Hogerheijde}, {Wilner}, \&
  {Qi}}]{panic09}
{Pani{\'c}}, O., {Hogerheijde}, M.~R., {Wilner}, D., \& {Qi}, C. 2009, \aap,
  501, 269

\bibitem[{{P{\'e}rez} {et~al.}(2014){P{\'e}rez}, {Isella}, {Carpenter}, \&
  {Chandler}}]{perez14}
{P{\'e}rez}, L.~M., {Isella}, A., {Carpenter}, J.~M., \& {Chandler}, C.~J.
  2014, \apjl, 783, L13

\bibitem[{{P{\'e}rez} {et~al.}(2012){P{\'e}rez}, {Carpenter}, {Chandler},
  {Isella}, {Andrews}, {Ricci}, {Calvet}, {Corder}, {Deller}, {Dullemond},
  {Greaves}, {Harris}, {Henning}, {Kwon}, {Lazio}, {Linz}, {Mundy}, {Sargent},
  {Storm}, {Testi}, \& {Wilner}}]{perez12}
{P{\'e}rez}, L.~M., {et~al.} 2012, \apjl, 760, L17

\bibitem[{{Pi{\'e}tu} {et~al.}(2006){Pi{\'e}tu}, {Dutrey}, {Guilloteau},
  {Chapillon}, \& {Pety}}]{pietu06}
{Pi{\'e}tu}, V., {Dutrey}, A., {Guilloteau}, S., {Chapillon}, E., \& {Pety}, J.
  2006, \aap, 460, L43

\bibitem[{{Pi{\'e}tu} {et~al.}(2014){Pi{\'e}tu}, {Guilloteau}, {Di Folco},
  {Dutrey}, \& {Boehler}}]{pietu14}
{Pi{\'e}tu}, V., {Guilloteau}, S., {Di Folco}, E., {Dutrey}, A., \& {Boehler},
  Y. 2014, \aap, 564, A95

\bibitem[{{Pineda} {et~al.}(2014){Pineda}, {Quanz}, {Meru}, {Mulders}, {Meyer},
  {Pani{\'c}}, \& {Avenhaus}}]{pineda14}
{Pineda}, J.~E., {Quanz}, S.~P., {Meru}, F., {Mulders}, G.~D., {Meyer}, M.~R.,
  {Pani{\'c}}, O., \& {Avenhaus}, H. 2014, \apjl, 788, L34

\bibitem[{{Pinilla} {et~al.}(2012){Pinilla}, {Birnstiel}, {Ricci}, {Dullemond},
  {Uribe}, {Testi}, \& {Natta}}]{pinilla12}
{Pinilla}, P., {Birnstiel}, T., {Ricci}, L., {Dullemond}, C.~P., {Uribe},
  A.~L., {Testi}, L., \& {Natta}, A. 2012, \aap, 538, A114

\bibitem[{{Rebull} {et~al.}(2010){Rebull}, {Padgett}, {McCabe}, {Hillenbrand},
  {Stapelfeldt}, {Noriega-Crespo}, {Carey}, {Brooke}, {Huard}, {Terebey},
  {Audard}, {Monin}, {Fukagawa}, {G{\"u}del}, {Knapp}, {Menard}, {Allen},
  {Angione}, {Baldovin-Saavedra}, {Bouvier}, {Briggs}, {Dougados}, {Evans},
  {Flagey}, {Guieu}, {Grosso}, {Glauser}, {Harvey}, {Hines}, {Latter},
  {Skinner}, {Strom}, {Tromp}, \& {Wolf}}]{rebull10}
{Rebull}, L.~M., {et~al.} 2010, \apjs, 186, 259

\bibitem[{{Reggiani} {et~al.}(2014){Reggiani}, {Quanz}, {Meyer}, {Pueyo},
  {Absil}, {Amara}, {Anglada}, {Avenhaus}, {Girard}, {Carrasco Gonzalez},
  {Graham}, {Mawet}, {Meru}, {Milli}, {Osorio}, {Wolff}, \&
  {Torrelles}}]{reggiani14}
{Reggiani}, M., {et~al.} 2014, \apjl, 792, L23

\bibitem[{{Rigliaco} {et~al.}(2012){Rigliaco}, {Natta}, {Testi}, {Randich},
  {Alcal{\`a}}, {Covino}, \& {Stelzer}}]{rigliaco12}
{Rigliaco}, E., {Natta}, A., {Testi}, L., {Randich}, S., {Alcal{\`a}}, J.~M.,
  {Covino}, E., \& {Stelzer}, B. 2012, \aap, 548, A56

\bibitem[{{Rosenfeld} {et~al.}(2013){Rosenfeld}, {Andrews}, {Wilner},
  {Kastner}, \& {McClure}}]{rosenfeld13}
{Rosenfeld}, K.~A., {Andrews}, S.~M., {Wilner}, D.~J., {Kastner}, J.~H., \&
  {McClure}, M.~K. 2013, \apj, 775, 136

\bibitem[{{Siess} {et~al.}(2000){Siess}, {Dufour}, \& {Forestini}}]{siess00}
{Siess}, L., {Dufour}, E., \& {Forestini}, M. 2000, \aap, 358, 593

\bibitem[{{Skrutskie} {et~al.}(1990){Skrutskie}, {Dutkevitch}, {Strom},
  {Edwards}, {Strom}, \& {Shure}}]{skrutskie90}
{Skrutskie}, M.~F., {Dutkevitch}, D., {Strom}, S.~E., {Edwards}, S., {Strom},
  K.~M., \& {Shure}, M.~A. 1990, \aj, 99, 1187

\bibitem[{{Skrutskie} {et~al.}(2006){Skrutskie}, {Cutri}, {Stiening},
  {Weinberg}, {Schneider}, {Carpenter}, {Beichman}, {Capps}, {Chester},
  {Elias}, {Huchra}, {Liebert}, {Lonsdale}, {Monet}, {Price}, {Seitzer},
  {Jarrett}, {Kirkpatrick}, {Gizis}, {Howard}, {Evans}, {Fowler}, {Fullmer},
  {Hurt}, {Light}, {Kopan}, {Marsh}, {McCallon}, {Tam}, {Van Dyk}, \&
  {Wheelock}}]{skrutskie06}
{Skrutskie}, M.~F., {et~al.} 2006, \aj, 131, 1163

\bibitem[{{Strom} {et~al.}(1989){Strom}, {Strom}, {Edwards}, {Cabrit}, \&
  {Skrutskie}}]{strom89}
{Strom}, K.~M., {Strom}, S.~E., {Edwards}, S., {Cabrit}, S., \& {Skrutskie},
  M.~F. 1989, \aj, 97, 1451

\bibitem[{{Uchida} \& {Shibata}(1984)}]{uchida84}
{Uchida}, Y., \& {Shibata}, K. 1984, \pasj, 36, 105

\bibitem[{{van der Marel} {et~al.}(2013){van der Marel}, {van Dishoeck},
  {Bruderer}, {Birnstiel}, {Pinilla}, {Dullemond}, {van Kempen}, {Schmalzl},
  {Brown}, {Herczeg}, {Matthews}, \& {Geers}}]{vandermarel13}
{van der Marel}, N., {et~al.} 2013, Science, 340, 1199

\bibitem[{{Weidenschilling}(1977)}]{weidenschilling77}
{Weidenschilling}, S.~J. 1977, \mnras, 180, 57

\bibitem[{{White} \& {Basri}(2003)}]{white03}
{White}, R.~J., \& {Basri}, G. 2003, \apj, 582, 1109

\bibitem[{{White} \& {Ghez}(2001)}]{white01}
{White}, R.~J., \& {Ghez}, A.~M. 2001, \apj, 556, 265

\bibitem[{{Wright} {et~al.}(2010){Wright}, {Eisenhardt}, {Mainzer}, {Ressler},
  {Cutri}, {Jarrett}, {Kirkpatrick}, {Padgett}, {McMillan}, {Skrutskie},
  {Stanford}, {Cohen}, {Walker}, {Mather}, {Leisawitz}, {Gautier}, {McLean},
  {Benford}, {Lonsdale}, {Blain}, {Mendez}, {Irace}, {Duval}, {Liu}, {Royer},
  {Heinrichsen}, {Howard}, {Shannon}, {Kendall}, {Walsh}, {Larsen}, {Cardon},
  {Schick}, {Schwalm}, {Abid}, {Fabinsky}, {Naes}, \& {Tsai}}]{wright10}
{Wright}, E.~L., {et~al.} 2010, \aj, 140, 1868

\bibitem[{{Zacharias} {et~al.}(2009){Zacharias}, {Finch}, {Girard}, {Hambly},
  {Wycoff}, {Zacharias}, {Castillo}, {Corbin}, {Divittorio}, {Dutta}, {Gaume},
  {Gauss}, {Germain}, {Hall}, {Hartkopf}, {Hsu}, {Holdenried}, {Makarov},
  {Martinez}, {Mason}, {Monet}, {Rafferty}, {Rhodes}, {Siemers}, {Smith},
  {Tilleman}, {Urban}, {Wieder}, {Winter}, \& {Young}}]{zacharias09}
{Zacharias}, N., {et~al.} 2009, VizieR Online Data Catalog, 1315, 0

\bibitem[{{Zhang} {et~al.}(2014){Zhang}, {Isella}, {Carpenter}, \&
  {Blake}}]{zhang14}
{Zhang}, K., {Isella}, A., {Carpenter}, J.~M., \& {Blake}, G.~A. 2014, \apj,
  791, 42

\bibitem[{{Zhu} {et~al.}(2012){Zhu}, {Nelson}, {Dong}, {Espaillat}, \&
  {Hartmann}}]{zhu12}
{Zhu}, Z., {Nelson}, R.~P., {Dong}, R., {Espaillat}, C., \& {Hartmann}, L.
  2012, \apj, 755, 6

\bibitem[{{Zhu} {et~al.}(2011){Zhu}, {Nelson}, {Hartmann}, {Espaillat}, \&
  {Calvet}}]{zhu11}
{Zhu}, Z., {Nelson}, R.~P., {Hartmann}, L., {Espaillat}, C., \& {Calvet}, N.
  2011, \apj, 729, 47

\end{thebibliography}

\end{document}